\documentclass[12pt]{article}
\usepackage{pst-plot,epsf}
\newcommand{\beq}{\begin{equation}}
\newcommand{\eeq}{\end{equation}}
\newcommand{\bea}{\begin{eqnarray}}
\newcommand{\eea}{\end{eqnarray}}

\newcommand{\epm}{e^+e^-}

\newcommand{\ra}{\rightarrow}

\newcommand{\eetbmn}{e^+ e^- \ra t \bar{b} \mu^- \bar{\nu}_{\mu}}
\newcommand{\eetben}{e^+ e^- \ra t \bar{b} e^- \bar{\nu}_{e}}

\newcommand{\tbmn}{t \bar{b} \mu^- \bar{\nu}_{\mu}}
\newcommand{\tbtn}{t \bar{b} \tau^- \bar{\nu}_{\tau}}
\newcommand{\tbdu}{t \bar{b} d \bar{u}}
\newcommand{\tbsc}{t \bar{b} s \bar{c}}
\begin{document}
\thispagestyle{empty}
\begin{flushright}
TP-USl-01/01\\
hep-ph/0102253\\
February 2001\\
\vspace*{1.5cm}
\end{flushright}
\begin{center}
{\LARGE\bf Top quark production in $e^+ e^-$ 
           annihilation\footnote{Work supported
           in part by the Polish State Committee for Scientific Research
           (KBN) under contract No. 2~P03B~004~18.}}\\
\vspace*{2cm}
Adam Biernacik, Katarzyna Cie\'ckiewicz \\[2mm]
and \\[2mm]
Karol Ko\l odziej\footnote{E-mail: kolodzie@us.edu.pl}\\[1cm]
{\small\it
Institute of Physics, University of Silesia\\ 
ul. Uniwersytecka 4, PL-40007 Katowice, Poland}\\
\vspace*{3.5cm}
{\bf Abstract}\\
\end{center}
We analyze the four-fermion reactions $e^+ e^- \ra 4f$ containing a single
top quark and three other fermions, a possible decay product of the resonant
anti-top quark, in the final state. This allows us to estimate the contribution
of the nonresonant Feynman graphs and effects related to the off mass shell 
production and decay of the top quark. We test the sensitivity of the total 
cross section at centre of mass energies in the $t\bar{t}$ threshold region 
and far above it to the variation of the top quark width.
We perform calculation in an arbitrary linear gauge 
in the framework of the Standard Model and discuss an important issue
of gauge symmetry violation by the constant top quark width.
\vfill
\newpage
\section{Introduction}
Production of the top quark in $\epm$ annihilation is an issue which
has attracted a lot of interest of both experimenters and theorists
for the past decade. In the present decade the interest will be certainly 
growing up in the prospect of new high energy $\epm$ colliders gradually
coming up from the stage of general discussion to a detailed planning and 
hopefully to the stage of construction and successful operation \cite{Zerwas}.
As the $\epm$ machines operate in a very clean experimental environment
they provide a unique possibility of precise measurement of the top
quark physical properties which, as many expect, may go beyond the 
standard model (SM) and give decisive hints for the development of new 
physical ideas. In order to disentangle the possible effects of new physics 
from the standard physics it is crucial to know the SM predictions as 
precisely as possible.

It has taken a lot of effort to obtain precise SM predictions
for the top pair production in the threshold region.
A substantial improvement of convergence of the perturbation series
has been achieved by computing the next-to-next-to-leading order 
corrections to the top quark pair production cross section
near threshold \cite{Sumino}
and understanding the renormalon cancellation mechanism \cite{Hoang}. 
The $\mathcal{O}(\alpha\alpha_s)$ corrections to the top decay with
into a $W$ boson and a $b$ quark are also known \cite{Jezabek}.

In the present note, we concentrate on the effects
related to the fact that one of the quarks in the $t\bar{t}$ pair may
be produced off mass shell. In particular, we let the anti-top quark 
$\bar{t}$ decay into a final state possible in the framework of the SM,
i.e., we consider reactions
\beq
\label{reactions}
         \epm \ra t \bar{b} f \bar{f'},
\eeq
where $f = e^-,\mu^-,\tau^-,d,s$ and 
$f' = \nu_{e},\nu_{\mu},\nu_{\tau},u,c$, respectively, 
taking into account the complete set of the Feynman graphs which contribute
to the specific final state at the tree level.
We pay attention to the very important issue of gauge symmetry 
breaking caused by the nonzero constant widths of unstable particles, in 
particular that of the top quark, which are kept as free parameters.
We also test the sensitivity of the total cross sections of 
(\ref{reactions}) to the variation of the top width. 
The deviation of the top width from its SM value may indicate the new physics.
In order to test the 
reliability of our results, we perform the calculation in the arbitrary 
linear $R_{\xi}$ gauge. 
We neglect radiative corrections, the correct treatment of 
which demands an extra effort and is beyond the scope of the present work.

We describe basics of the calculation in the next section. Our results
are presented and discussed in Section 3 and, finally, in Section 4,
we give our concluding remarks.
\section{Calculation in arbitrary linear gauge}
The calculation of the necessary matrix elements relies on the method
proposed in Ref.~\cite{KZ} and further developed in Ref.~\cite{JK}.
As in Ref.~\cite{JK}, fermion masses are kept nonzero both in
the matrix elements and in the kinematics. Among others, this has
an advantage that the Higgs boson effects can be incorporated 
consistently and a pole related to the photon exchange in the $t$-channel 
can be handled better than in the massless fermion case.

In order to estimate the gauge symmetry violation effects related to
the nonzero widths of unstable particles, we perform the calculation
in two different schemes: the {\em `fixed widths scheme'} (FWS) 
and in the so called {\em `complex-mass scheme'} (CMS) of 
Ref.~\cite{DDRW1}. Both schemes introduce the constant particle widths 
through the complex mass parameters:
\beq
\label{cmass}
M_V^2=m_V^2-im_V\Gamma_V,  \quad V=W, Z, H, \qquad {\rm and}
\qquad M_t=m_t-i\Gamma_t/2,
\eeq
which replace masses in the corresponding propagators. The coupling 
constants are given in terms of the electric charge and
electroweak mixing parameter $\sin^2\theta_W$. In FWS, the electroweak
mixing parameter is kept real, i.e., it is given by
\beq
\label{rsw2}
\sin^2\theta_W=1-m_W^2/m_Z^2,
\eeq
where $m_W$ and $m_Z$ are physical masses of the $W^{\pm}$ and $Z^0$ boson.
In CMS, $\sin^2\theta_W$ is given in terms of the complex masses
$M_W$ and $M_Z$ of Eq.~(\ref{cmass}) by
\beq
\label{csw2}
\sin^2\theta_W=1-M_W^2/M_Z^2,
\eeq
thus it is a complex number.
The CMS has the advantage that it preserves the Ward identities \cite{DDRW1},
provided all the fermion widths are zero. We have tested the gauge invariance
numerically for all the reactions (\ref{reactions}) under the assumption 
of the zero top quark widths. With the nonzero top width introduced
in the Feynman propagator of the top quark
\beq
\label{tprop}
i\;S_t^F(p)=i \;
{{/\!\!\!p + M_t}\over{p^2 - M_t^2}},
\eeq
the Ward identities are not satisfied any more and hence the gauge
symmetry is violated.

There has been a lot of discussion of the gauge invariance issue in 
literature of the past few years \cite{finwid}. The best way to solve 
the problem, however involved and complicated it might be, is to work 
in the {\em `fermion-loop
scheme'} (FLS), which resumes higher order effects coming
from one-loop fermion contributions to bosonic propagators together 
with parts of the vertices necessary
for keeping the corrections gauge invariant. Unfortunately, a similar 
Dyson resummation has not been worked out in a detailed way for a propagator
of an unstable fermion up to now.
Therefore, the substitution of Eq.~(\ref{cmass}) has no full theoretical
justification in the framework of quantum field theory. However,
a comparison of numerical results obtained in the schemes using fixed widths
in both the $s$- and $t$-channel gauge boson propagators with those derived
within FLS, which show no numerically relevant discrepancies for
several four-fermion reactions and the corresponding bremsstrahlung 
processes, speaks in favour of the simplified approach based on
substitution of Eq.~(\ref{cmass}) in each propagator of unstable particle.
Therefore, in the following we will restrict ourselves to this
simplified approach.

In order to quantitatively estimate gauge symmetry violation effects
induced by the substitution of Eq.~(\ref{cmass}), we perform calculation 
of the matrix elements in the linear gauge with arbitrary real gauge 
parameters $\xi_V, V=\gamma, W, Z$. We then vary the parameters in
a very wide range from $\xi_V=1$ corresponding to the 'tHooft--Feynman 
gauge (FG) to $\xi_V=10^{16}$ which, in the double precision of Fortran
programming language, corresponds to the unitary gauge (UG). We would like to
stress at this point that we take into account contributions from the 
exchange of the would-be Goldstone bosons in the $R_{\xi}$ gauge, which
are absent in the unitary gauge.
If the change in the cross section induced by the change in gauge parameters
is smaller than the accuracy of the Monte Carlo integration we assume
that the gauge violation effects induced by the nonzero top width are 
numerically irrelevant and we may consider the corresponding results as 
trustworthy. On the other hand, if the results depend on the choice
of gauge parameters they are useless, but we may try to reduce the 
dependence by imposing cuts on the phase space integration.
This simple prescription, however doubtful from the purely
theoretical point of view might it be, allows one to treat the particle 
widths as independent parameters and test the reliability of the numerical
results in an efficient way.

The phase space integration is performed numerically using a
multichannel Monte Carlo (MC) approach and the integration routine
{\tt VEGAS} \cite{vegas}. The 7 dimensional phase space element of 
reaction (\ref{reactions}) is parametrized in a few different ways
in order to account for the most relevant peaks of the matrix elements:
the $\sim 1/t$ pole caused by the $t$-channel photon-exchange,
the Breit--Wigner shape of the $W^{\pm}$ and $Z^0$ resonances,
the $\sim 1/s$ behavior of a light fermion pair production,
and the $\sim 1/t$ pole due to the the neutrino exchange at
the same time.

\section{Numerical results}
In this section, we will present numerical results for all the
four-fermion channels of reaction (\ref{reactions}) possible in the SM.

We define the SM physical parameters in terms of the gauge boson masses
and widths, the top mass and the Fermi coupling constant. We take the
actual values of the parameters from Ref.~\cite{PDG}:\\[4mm]
\centerline{
$m_W=80.419\; {\rm GeV}, \quad \Gamma_W=2.12\; {\rm GeV}, \qquad
m_Z=91.1882\; {\rm GeV}, \quad \Gamma_Z=2.4952\; {\rm GeV}$,}
\beq
\label{params}
m_t=174.3\; {\rm GeV}, \quad G_{\mu}=1.16639 \times 10^{-5}\;{\rm GeV}^{-2}.
\eeq
We assume the Higgs boson mass of $m_H=115$ GeV and, if not stated otherwise, 
the top quark width of $\Gamma_t=1.5$ GeV.

For the sake of definiteness we also list other fermion masses we 
use in the calculation \cite{PDG}:\\[4mm]
\centerline{
$m_e=0.510998902\; {\rm MeV}, \quad m_{\mu}=105.658357\; {\rm MeV},\quad
m_{\tau}=1777.03\; {\rm MeV}$,}
\beq
m_u=5\; {\rm MeV}, \quad m_d=9\; {\rm MeV}, \quad m_s=150\; {\rm MeV}, \quad
m_c=1.3\; {\rm GeV}, \quad m_b=4.4\; {\rm GeV}.
\eeq
We neglect the Cabibo--Kobayashi--Maskawa mixing, i.e., we assume the
CKM matrix to be a unit matrix.

The fine structure constant is calculated from
\beq
\label{alphaw}
\alpha_W=\sqrt{2} G_{\mu} m_W^2 \sin^2\theta_W/\pi
\eeq
with the real electroweak mixing parameters of Eq. (\ref{rsw2}) in the
both schemes FWS and CMS.

Except for the check of gauge invariance discussed in the previous
section, we perform a few other checks. Our results reproduce those
of Ref. \cite{JK} for a top mass smaller than $m_W$ and the zero top width.
The corresponding matrix elements in the absence of the Higgs boson
exchange has been checked against {\tt MADGRAPH} \cite{MADGRAPH}.
The phase space generation routine for particles of large masses
has been written in two independent ways.

In Table~1, we show the results for the cross sections of $\eetbmn$ at 
different centre of mass energies obtained in different schemes and gauges: 
the complex-mass scheme (CMS), fixed width scheme (FWS), unitary gauge (UG)
and Feynman gauge (FG). We have integrated over the full four particle 
phase space without any cuts. We can see that the results hardly depend on 
the gauge choice both in the CMS and FWS. Actually, they nicely agree
with each other within one standard deviation of the MC integration.

In Table~2, we present the results for $\eetben$ obtained in the CMS
and in two different gauges, UG and FG. In order to reduce the
dependence on gauge choice induced by the nonzero top width, we have imposed 
a cut on the electron angle with respect to the beam 
$\theta(e^-,{\rm beam})$. Again there is rather small dependence on the cut
for the energies presented in Table~2.
From a comparison with the corresponding numbers of Table~1, we can infer
that the $t$-channel Feynman graphs of reaction $\eetben$ do not contribute
much to the total cross section in the presence of the cut on the final
electron angle.
When we reduce the cut further so that 
the denominator of the $t$-channel photon propagator becomes
of the order of the electron mass squared the dependence on gauge becomes
substantial and results are meaningless.

\begin{center}
Table~1: Cross sections in fb of $\eetbmn$ at different centre of mass energies
         in different schemes, CMS and FWS, and gauges, UG and FG.
         The numbers in parenthesis show the uncertainty of the last 
         decimals.\\[5mm]
\begin{tabular}{|c|c|c|c|c|}
\hline
\rule{0mm}{7mm} $\sqrt{s}$ (GeV) & ${\sigma}_{CMS}^{UG}$ 
      & ${\sigma}_{CMS}^{FG}$
                           & ${\sigma}_{FWS}^{UG}$ 
                               & ${\sigma}_{FWS}^{FG}$ \\[2mm]
\hline
\rule{0mm}{7mm} 190 & $ 2.6174(7) \times 10^{-8}$ 
         & $ 2.6174(7) \times 10^{-8} $
                 & $ 2.6185(7) \times 10^{-8}$ & $ 2.6185(7) \times 10^{-8}$\\
 340 & 0.7837(4) & 0.7837(4) & 0.7839(4) & 0.7840(4) \\
 360 & 41.27(10)   & 41.27(10)   & 41.28(10)   & 41.29(10)   \\
 500 & 60.06(13)   & 60.04(13)   & 59.75(30)   & 59.90(29)   \\
2000 & 5.59(3)   & 5.56(3)   & 5.51(7)   & 5.51(8)   \\[2mm]
\hline
\end{tabular}
\end{center}

\begin{center}
Table~2: Cross sections in fb of $\eetben$ in the CMS in two different 
         gauges, UG and FG, and for two different 
         cuts on the electron angle with respect to the beam.\\[5mm]
\begin{tabular}{|c|c|c|c|c|}
\hline
\rule{0mm}{7mm} $\sqrt{s}$ 
 & \multicolumn{2}{c|}{$ 5^0 < \theta(e^-,{\rm beam}) < 175^0 $}
      & \multicolumn{2}{c|}{$ 1^0 < \theta(e^-,{\rm beam}) < 179^0 $}\\[2mm]
\cline{2-5}
\rule{0mm}{7mm} (GeV) & ${\sigma}_{CMS}^{UG}$ & ${\sigma}_{CMS}^{FG}$
                  & ${\sigma}_{CMS}^{UG}$ & ${\sigma}_{CMS}^{FG}$ \\[2mm]
\hline
\rule{0mm}{7mm} 190 
      & $ 0.6607(4) \times 10^{-5}$ & $ 0.6607(4) \times 10^{-5}$ 
          & $ 0.10520(4) \times 10^{-4} $ & $ 0.10531(4) \times 10^{-4} $ \\
 340 & 0.7993(4) & 0.7993(4) & 0.8251(4) & 0.8253(4) \\
 360 & 41.21(11)   & 41.20(11)   & 41.32(8) & 41.32(8) \\
 500 & 59.78(15)   & 59.75(15)   & 60.16(15) & 60.19(15) \\
2000 & 6.81(3)   & 6.82(3)   & 7.97(3) & 8.00(3) \\[2mm]
\hline
\end{tabular}
\end{center}

The results for the channels of reaction (\ref{reactions}) which do
not contain electron in the final state are shown in Table~3. They
were obtained in the CMS and unitary gauge.

\vfill

\newpage

\begin{center}
Table~3: Cross sections in the CMS in fb of different channels of reaction
         (\ref{reactions}) not containing a final state electron.\\[5mm]
\begin{tabular}{|l|c|c|c|c|c|}
\hline
Channel of &  \multicolumn{5}{|c|}{\rule{0mm}{7mm}$\sqrt{s}$ (GeV)} \\[2mm]
\cline{2-6}
reaction (\ref{reactions}) & \rule{0mm}{7mm} 190 & 340 & 360 & 500 
                         & 2000 \\[2mm]
\hline
$e^+e^- \ra \tbmn$ & \rule{0mm}{7mm} $2.6174(7) \times 10^{-8}$
                        & 0.7837(4) & 41.3(1) & 59.8(3) & 5.42(7) \\[2mm]
$e^+e^- \ra \tbtn$ & $1.9331(4) \times 10^{-8}$& 0.7831(4) & 41.2(1) 
                        & 59.6(3) & 5.47(7) \\[2mm]
$e^+e^- \ra \tbdu$ & $7.880(2) \times 10^{-8}$ & 2.351(1)  & 123.8(3)
                        & 179.9(9)& 16.3(2) \\[2mm]
$e^+e^- \ra \tbsc$ & $6.616(2) \times 10^{-8}$ & 2.350(1)  & 123.8(3)
                   & 178.9(9) & 16.7(2) \\[2mm]
\hline
\end{tabular}
\end{center}

In Fig.~1, we show the energy dependence of the total cross section
of $e^+e^- \ra t\bar{b}\mu^-\bar{\nu}_{\mu}$ calculated with the
complete set of Feynman graphs and the approximate
cross section $e^+e^- \ra t\bar{t} \ra t\bar{b}\mu^-\bar{\nu}_{\mu}$.
The latter has been obtained by multiplying the on shell top pair
production cross section by the corresponding three body top decay width
\beq
\label{approx}
\sigma(e^+e^- \ra t\bar{t} \ra t\bar{b}\mu^-\bar{\nu}_{\mu})=
\sigma(e^+e^- \ra t\bar{t}) \; \Gamma(\bar{t} \ra \bar{b}\mu^-\bar{\nu}_{\mu}).
\eeq
We have taken over the SM part of the analytic formula for the width
$\Gamma(\bar{t} \ra \bar{b}\mu^-\bar{\nu}_{\mu})$ with massless
final state fermions from Ref.~\cite{Boos}. In the calculation of 
$\sigma(e^+e^- \ra t\bar{b}\mu^-\bar{\nu}_{\mu})$ we have used the FWS 
scheme and neglected the Higgs boson contribution.
We see that Eq.~(\ref{approx})
approximates the complete tree level calculation well not only
just above the threshold, but also for higher centre of mass energies.
The relative difference between the both results is 3.5\% at 360 GeV,
1.3\% at 500 GeV and - 5.0\% at 800 GeV. The nice agreement is somewhat 
amazing as except for one Feynman graph which contain a resonant top
propagator there are nine other nonresonant graphs 
which contribute to $e^+e^- \ra t\bar{b}\mu^-\bar{\nu}_{\mu}$
in the unitary gauge.

The explanation of this fact can easily be found if one looks at Fig.~2
where we have plotted, against the centre of mass energy, the cross section 
of Eq.~(\ref{approx}) and another approximated cross section obtained by 
integrating over full four particle phase space the squared matrix 
element containing only the top resonant Feynman graph.
The small discrepancy between the two curves in Fig.~2 is a measure of spin
correlations and off shellness of the $\bar{t}$ quark.

We illustrate the dependence of the total cross section of $\eetbmn$ 
on the top quark width $\Gamma_t$ in Table~4. We see that the cross section at
$\sqrt{s} =360$ GeV, i.e. just above the threshold is almost exactly
proportional to $1/\Gamma_t$. This kind of dependence holds also at
$\sqrt{s} =500$ GeV, which is already much above the threshold and it 
survives almost unaltered at $\sqrt{s} = 2$ TeV. It means that the cross 
section of $\eetbmn$ is well approximated by the resonant $\bar{t}$ production 
and its subsequent decay. This kind of dependence offers a new way of 
measurement of the top quark width alternative to the measurement based 
on the shape of the $t\bar{t}$ threshold \cite{Comas}.

\rput(7,-6){\scalebox{0.8 0.8}{\epsfbox{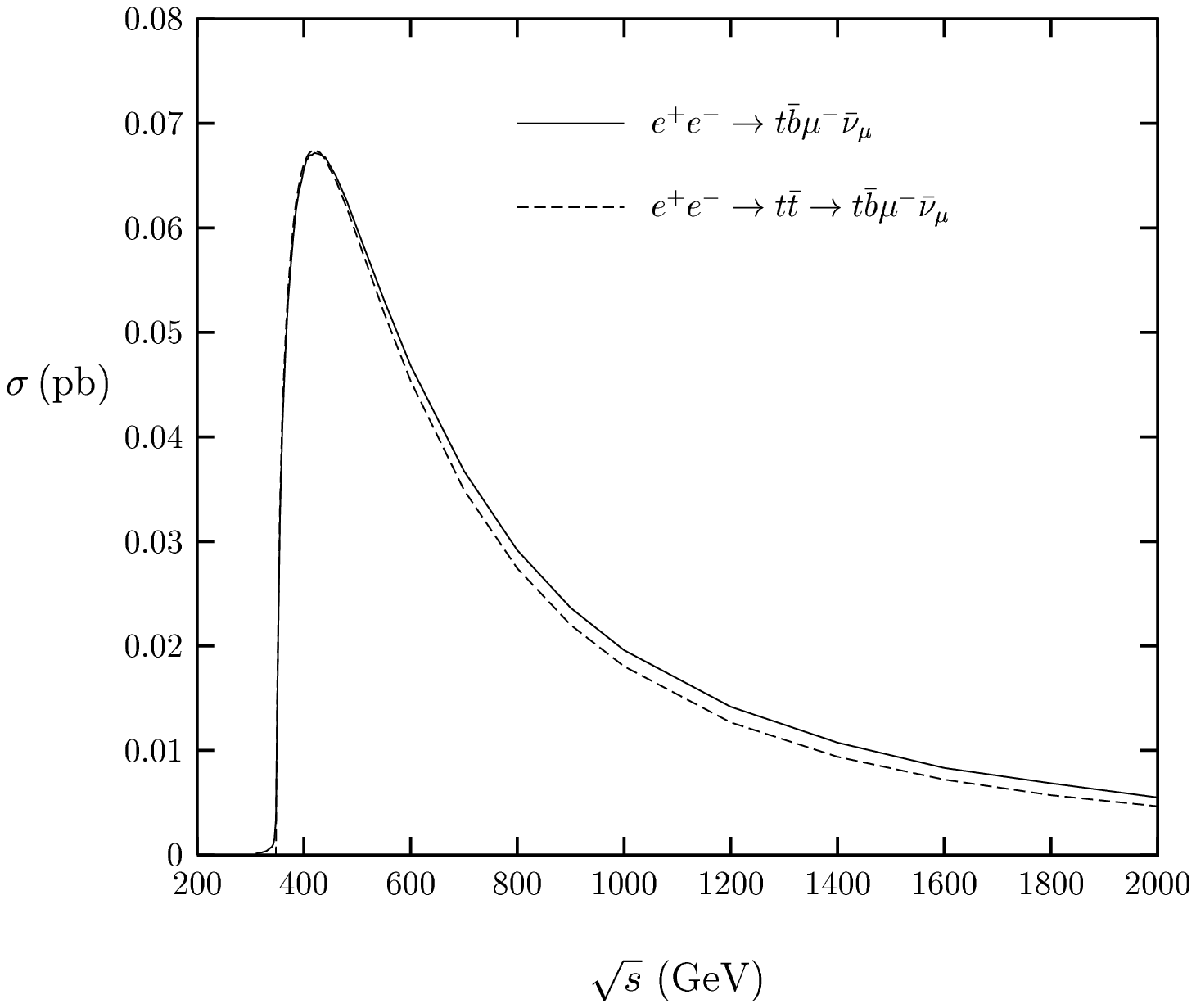}}}

\vspace*{9.0cm}

\bigskip
\bigskip
\begin{center}
{\small Figure~1. The energy dependence of the total cross sections
of $e^+e^- \ra t\bar{b}\mu^-\bar{\nu}_{\mu}$
and $e^+e^- \ra t\bar{t} \ra t\bar{b}\mu^-\bar{\nu}_{\mu}$.}
\end{center}

\rput(7,-6){\scalebox{0.8 0.8}{\epsfbox{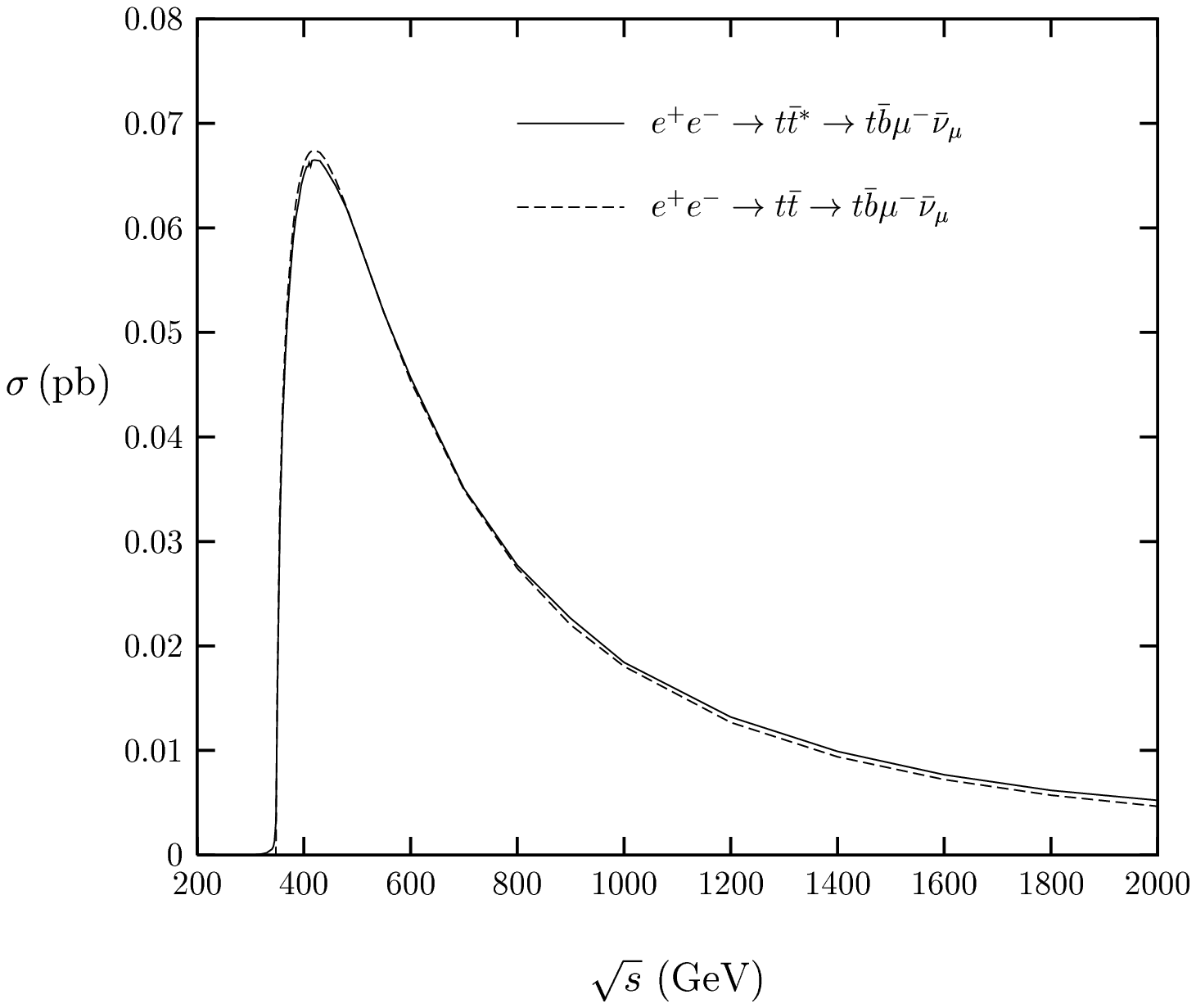}}}

\vspace*{9.0cm}

\bigskip
\bigskip
\bigskip

\begin{center}
{\small Figure~2. The energy dependence of the total cross sections
of $e^+e^- \ra t\bar{t}^* \ra t\bar{b}\mu^-\bar{\nu}_{\mu}$
and $e^+e^- \ra t\bar{t} \ra t\bar{b}\mu^-\bar{\nu}_{\mu}$.}
\end{center}

\begin{center}
Table~4: Cross sections in fb of $\eetbmn$ for different values of the
         top quark width. The calculation has been performed in the CMS 
         and UG.\\[5mm]
\begin{tabular}{|c|c|c|c|}
\hline
$\sqrt{s}$  &  \multicolumn{3}{|c|}{\rule{0mm}{7mm}$\Gamma_t$ (GeV)} \\[2mm]
\cline{2-4}
\rule{0mm}{7mm} (GeV) & 1.5 & 1.6 & 1.7 \\[2mm]
\hline
\rule{0mm}{7mm} 
 190 & $ 2.6174(7) \times 10^{-8}$ & $ 2.6186(4) \times 10^{-8}$ 
                                   & $ 2.6186(4) \times 10^{-8}$ \\[2mm]
 340 & 0.7837(4) & 0.7832(3) & 0.7830(3) \\[2mm]
 360 & 41.27(10) & 38.65(7)  & 36.31(6) \\[2mm]
 500 & 60.06(13) & 56.37(13) & 53.13(12) \\[2mm]
2000 & 5.59(3)   &  5.30(2)  &  5.09(2) \\[2mm]
\hline
\end{tabular}
\end{center}

\section{Summary and Outlook}
We have analyzed the top quark production in $\epm$ annihilation
at a new high luminosity linear collider like TESLA. We have 
estimated the contribution of the 
nonresonant Feynman graphs and effects related to the off mass shell 
production and decay of one of the top quarks. Those effects are typically
of the order of a few per cent. Therefore one should take them into account
in the analysis of the future data. We have shown that the cross
section of reaction (\ref{reactions}) is dominated by the
resonant $\bar{t}$ production and its subsequent decay not only
at centre of mass energies in the $t\bar{t}$ threshold region but
also far above it. 
We have tested the sensitivity of the total cross sections 
to the variation of the top quark width and confirmed expected
proportionality to $1/\Gamma_t$ over a very wide energy range
beginning from the threshold. This kind of dependence offers an alternative
way of measurement of the top quark width.
By performing calculation  in an arbitrary linear gauge 
in the framework of the Standard Model we have been able to address an 
important issue of gauge symmetry violation by the constant top quark width.

It would be desirable to consider the effects related to the off shell
production of the second quark of the $t\bar{t}$ pair and to include 
leading radiative corrections in the analysis of the future data.

\bigskip
{\bf Acknowledgment}
The authors are thankful to Fred Jegerlehner and DESY Zeuthen for kind
hospitality during their visits at the Institute.

\bigskip

\end{document}